\documentclass[12pt, titlepage, reqno]{article}
 \usepackage[super]{natbib}
 
 \usepackage{amssymb, latexsym}
 \usepackage{amsmath} 
 \usepackage{amsthm}
 \usepackage{bm}
 \usepackage[mathscr]{eucal}
 \usepackage{enumerate}

 \renewcommand{\section}[1]{\medskip \addtocounter{section}{1}\raggedright 
     \textbf{\Roman{section}. \ #1}\medskip \setcounter{subsection}{0}
    \setlength{\parindent}{5ex}
 }
 \renewcommand{\subsection}[1]{\medskip \addtocounter{subsection}{1}\raggedright
    \textbf{\Alph{subsection}. \ #1} \medskip \setcounter{subsubsection}{0}\setlength{\parindent}{5ex}
}

 \theoremstyle{plain}
\usepackage{graphicx} %  to include figures

\usepackage{epstopdf} % to include eps
\usepackage[font=footnotesize]{caption} % set the caption fontsize
 \usepackage[english]{babel}
 
 \newenvironment{sistema}%
{\left\lbrace\begin{array}{@{}l@{}}}%
{\end{array}\right.}

\begin{document}
  
     \begin{titlepage}

 \begin{center}

 \textbf{Hermite regularization of the Lattice Boltzmann Method}\\
 
 \textbf{for open source computational aeroacoustics}\\

 \vspace{10ex}

 F. Brogi
 \footnote{e-mail: fbrogi@inogs.it}$^{,}$\footnote{Also at the Department of Computer Science, University of Geneva, Route de Drize 7, CH-1227 Geneva, Switzerland.
  Now at: Istituto Nazionale di Oceanografia e di Geofisica Sperimentale (OGS) and Istituto Nazionale di Geofisica e Vulcanologia (INGV), Italy.}\\
 
Department of Earth Science, University of Geneva,

Rue des Maraîchers 13, CH-1205 Geneva, Switzerland.\\
 
  \vspace{4ex}
 
O. Malaspinas and B. Chopard \\
 
Department of Computer Science, University of Geneva,

Route de Drize 7, CH-1227 Geneva, Switzerland \\
  
  \vspace{4ex}
 
C. Bonadonna \\
 
Department of Earth Science, University of Geneva,

Rue des Maraîchers 13, CH-1205 Geneva, Switzerland \\

  \vspace{10ex}
 
 \today

 \end{center}

 \end{titlepage}

  \begin{abstract}
    The lattice Boltzmann method (LBM) is emerging as a powerful engineering tool for
    aeroacoustic computations. However, the LBM has been shown to present accuracy and
    stability issues in the medium-low Mach number range, that is of interest for aeroacoustic
    applications. Several solutions have been proposed but often are too computationally expensive,
    do not retain the simplicity and the advantages typical of the LBM, or are not described well
    enough to be usable by the community due to proprietary software policies. We propose to
    use an original regularized collision operator, based on the expansion in Hermite polynomials,
    that greatly improves the accuracy and stability of the LBM without altering significantly its algorithm.
    The regularized LBM can be easily coupled with both non-reflective boundary conditions and a multi-level
    grid strategy, essential ingredients for aeroacoustic simulations. Excellent agreement was found between
    our approach and both experimental and numerical data on two different benchmarks: the laminar, unsteady
    flow past a 2D cylinder and the 3D turbulent jet.  Finally, most of the aeroacoustic computations with LBM
    have been done with commercial softwares, while here the entire theoretical framework is implemented on top of an
    open source library (Palabos).
  \end{abstract}

\addtocounter{page}{2}

\section{INTRODUCTION}
 
\setlength{\parindent}{5ex}

  Numerical simulations represent a relatively low cost tool for predicting and improving our understanding of sound 
  production by turbulent flows and their interaction with solid boundaries. Model predictions based on Navier-Stokes (NS)
  solvers have been tested successfully on flow generated noise problems in many different applications with quite a wide
  range of Mach ($\mathrm{Ma}$) and Reynolds numbers ($\mathrm{Re}$).\citep{Colonius2004}${}^,$\citep{Wang2006}${}^,$\citep{Wagner2007} 
  More recently, there has been a growing interest in the Lattice Boltzmann method (LBM)
  to address aeroacoustic problems of practical
  relevance.\citep{Khorrami2016}$^,$\citep{Casalino2016}$^,$\citep{Casalino2014} This method instead of directly solving the
  the Navier-Stokes equations (NS) models fluids through the Boltzmann equation which describe the time evolution 
  of velocity density distribution function (see for instance Chopard \textit{et al.}\citep{Chopard2002}).
  The main differences with the NS solvers is the absence
  of non-linearity in the convection term and the fact that there is no need of an analytic description of the grid metrics.
  Such characteristics make the LBM, straightforward to code with a very good parallel efficiency, particularly adequate for solving complex fluid flows 
  and wave propagation with complicated geometry
  (e.g. Shin \textit{et al.},\cite{Shin2017} Zhu \textit{et al.}\cite{Zhu2016}).
  The simulation of  turbulent flows can be achieved, as for NS, with  Direct Numerical Simulation (LBM-DNS) 
  or Large Eddy Simulation (LBM-LES) approaches (e.g. Malaspinas and Sagaut\cite{Malaspinas2012}). The study of the flow over rectangular
  cavities,\citep{Ricot2001}
  the sound generated by a high-speed train in a tunnel\citep{Tsutahara2012} or the noise from a landing
  gear\citep{Noelting2010}${}^,$\citep{Casalino2014b}  
  are only few examples of successful LBM aeroacoustic applications. Interesting result have been obtained also for subsonic jet noise
  computations.\citep{Habibi2014}${}^,$\cite{Lew2010}${}^,$\cite{Lew2014} Lew \textit{et al.}\cite{Lew2010}${}^,$\cite{Lew2014} successfully simulate the noise generated for low and high subsonic
  jet including the nozzle geometry in the computational domain. The LBM turns out to be computationally more efficient with respect
  to the traditional NS-LES for low Mach number flow.
  Very promising are also the recent results obtained for the prediction of a complete turbofan engine
  noise\citep{Casalino2014}${}^,$\citep{Casalino2016} and full 
  aircraft airframe noise.\citep{Khorrami2014}
  All these examples clearly demonstrate the capability of the LBM-LES in solving aeroacoustic problems with complex geometries
  (in the subsonic regime) while remaining computationally efficient.
  In particular, most of these realistic applications have been solved thanks to the commercial software 
  PowerFLOW (http://www.exa.com/), developed by scientists that contributed significantly
  to the establishment of the LBM in automotive and aerospace engineering fields (e.g. Casalino \textit{et al.}\cite{Casalino2016}). Nonetheless, 
  the source code of this sophisticated software is not available to the end-user.
  In this work, we aim not only to present a complete LBM framework for dealing with aeroacoustic simulations but also to provide
  the aeroacoustic community with a computational tool entirely based on an open source library (PALABOS: http://www.palabos.org/). 
  
  Numerical schemes for computational aeroacoustics have to be capable not only to correctly resolve the pressure perturbations generated
  by the flow, which often span a wide range of frequencies, but also to propagate them with enough accuracy. This may become
  a difficult task to reach when considering the huge disparity in magnitude between the acoustic and the flow disturbances and
  the fact that acoustic waves may propagate with very low attenuation over long distances.
  Therefore, for accurate predictions several critical aspects of a numerical model have to be properly addressed. 
  
  Dissipation and dispersion errors induced by model discretization may be acceptable for the hydrodynamic fluctuations but can be unacceptable for the generation and propagation
  of the much smaller acoustic perturbation.
  Traditional NS solver for aeroacoustic computations are often based on high order low dissipative schemes, like finite-difference, with optimized 
  dispersive properties (e.g. Bogey and Bailly\cite{Bogey2004}). The most popular single relaxation time Lattice Boltzmann method (LBM-BGK) has been shown to have
  the low dissipative characteristics to properly capture the weak acoustic fluctuations and to be faster than the high order finite
  difference schemes for a given dispersion error.\citep{Marie2009}
  However the LBM-BGK, and more generally low dissipative schemes, generates instabilities which may induce the local divergence 
  of the computation and/or produce artificial noise in the form of 'numerical instability waves'. The latter are represented by high frequency
  spurious noise due to unadapted initial conditions, geometric singularities or large numerical approximations.\citep{Ricot2009} 
  Several solutions have been proposed to improve the stability of the LBM-BGK: imposing a local or global lower bound on the
  relaxation time,\citep{Li2004} increasing hyper-viscosities when non-uniform grids are used through Interpolation-Supplemented LBM, 
  \citep{Qian1997}${}^,$\citep{Fan2006}${}^,$\citep{Niu2004} enforcing the entropy H-theorem,
  \citep{Ansumali2002}${}^,$\citep{Brownlee2006}${}^,$\citep{Brownlee2007}${}^,$\citep{Brownlee2008}
  enhancing the bulk viscosity,\citep{Dellar2001} stability or dissipation/dispersion optimized multiple relaxation times schemes
  LBM-MRT,\citep{Lallemand2000}${}^,$\citep{Xu2011}${}^,$\citep{Dellar2014} selective viscosity filtering\citep{Ricot2009} and regularized procedure for LBM-BGK 
  scheme.\citep{Latt2006}
  Each of these methodologies has its own advantages and disadvantages and a complete comparison is still needed.
  Among the others the regularized procedure of Latt and Chopard\cite{Latt2006} emerges for its simplicity, 
  offering both increased accuracy and stability at very low cost.
  Based on this procedure but with a different approach, Malaspinas\cite{Malaspinas2015}
  has proposed an optimization of the LBM-BGK we adopted for our model. 
  While maintaining the simple framework of the classical LBM-BGK, the recursive and regularized LBM of
  Malaspinas\cite{Malaspinas2015} (LBM-rrBGK)
  provides a very stable and accurate scheme even at high Reynolds number and relatively high Mach number ($<0.6$).
  In particular, it has been shown to have very good dispersive and dissipation properties when analyzed
  by 2D Von-Neumann linear stability analysis.
  For now this approach is limited to the weakly compressible regime, but it appears to be extensible for 
  solving the dynamics of fully compressible and thermal flows. In fact, the LBM-rrBGK reduces dramatically
  the memory need which is one of the main issues to be resolved when the LBM is applied to this kind of flows.
  
  Regardless of the numerical method used to model fluid flows, open boundaries (infinite domains) cannot be dealt with easily but one has to
  truncate the domain with appropriate boundary conditions. It is crucial for aeroacoustics simulations, that
  the aerodynamic disturbances and the acoustic waves cross these artificial
  boundaries without producing significant reflections.
  Spurious reflected waves superimposed on the physical waves of interest, decrease the computational accuracy,
  and may influence the flow field driving the computed solution towards wrong (statistically) time-stationary state.\citep{Kam2007}
  In order to solve this problem, many different type of Non Reflecting Boundary Conditions (NRBC) has been developed in NS 
  (1D  non-reflecting characteristics-based boundary condition, sponge layer, the perfectly matched layer etc...).
  For the LBM instead, the formulation of NRBC is rather recent. Kam \textit{et al.}\cite{Kam2007} examines the applicability
  to the LBM of different types of NRBC commonly used in NS.
  Among four different schemes tested the 'sponge layer' performs the best. This type of NRBC, represents indeed an efficient
  way to prevent disturbances from reaching the domain boundaries.
  An artificial dissipation (damping force), only in a defined layer near the borders of the computational domain, damps
  all flow unsteadiness and force the flow towards a prescribed uniform solution. Recently, a rigorous formulation and optimization
  of the sponge layer has been given by  Xu and P. Sagaut.\cite{Xu2013} Analogue to the sponge layer, but based on a different formulation,
  is the perfectly matched layer (PML) developed by  Najafi-Yazdi and L. Mongeau.\cite{Najafi2012} Both the PML and the sponge
  layer have shown to have good stability properties and to be effective in damping the outgoing disturbances
  when tested on classical benchmark problems (e.g. the propagation of acoustic and vortex pulses).
   Let us note,
  that Lew \textit{et al.}\cite{Lew2010}${}^,$\cite{Lew2014} in his jet simulations avoid the use of NRBC by damping the outgoing
  disturbances with a very coarse grid near the boundaries. Requiring a big computational domain with respect to the smaller physical
  domain of interest, this strategy can be computationally demanding.

  Most fluid dynamics processes cover a wide range of spatial and temporal scales, which makes it difficult,
  if not impossible, to simulate all of them even when computations are performed on the most modern supercomputers.
  However, the smallest scales are often restricted to small regions of the computational domain (e.g near solid boundaries)
  and, therefore, as in more traditional fluid dynamics solvers,
  the definition of a grid that can be locally refined can partially overcome this problem. 
  In the Lattice Boltzmann community different solutions have been proposed for the grid refinement.
  In this work, we adopted a multi-domain refinement with a two-way coupling.\citep{Lagrava2012} In a parallel implementation,
  this approach helps to have better CPU performances and memory savings respect to the multi-grid refinement.
  \citep{Dupuis2003}${}^,$\citep{Freitas2006}${}^,$\citep{Rohde2006}${}^,$\citep{Yu2003}
  The communication of the information between grids at different resolution is of crucial importance for the accuracy, stability and
  computational performance. The stability of the scheme proposed by Lagrava \textit{et al.},\cite{Lagrava2012} is guarantee even
  at high Reynolds number by a
  filtering operation that decimate the information when passing from the fine to the coarse grid. Instead, when passing from the coarse
  to the fine grid, the reconstruction of the information is based on a local cubic interpolation scheme which maintains
  the order of accuracy of the LBM.   
  
  We aim to present a dedicated LBM model for aeroacoustic computations, which provides increased stability,
  accuracy and optimized computational performance thanks to the combination of a recursive and regularized LBM,
  (LBM-rrBGK)\citep{Malaspinas2015}
  an absorbing strategy as NRBC\citep{Xu2013} and multi-domain grid refinement with two way coupling.
  \citep{Lagrava2012} The model performances are tested by performing direct noise simulations for the two-dimensional circular cylinder
  in a uniform flow ($\mathrm{Ma} = 0.2, \mathrm{Re} = 150$) with a LBM-DNS and for a 3D subsonic turbulent jet with LBM-LES
  ($\mathrm{Ma} = 0.4, \mathrm{Re} = 6000$).
  For the latter, the model results will be compared against  experimental and numerical data from both NS
   and commercial LBM solvers (e.g. PowerFLOW). 
  Particular attention in the paper will be given to  provide a complete overview of all the of the theoretical and practical
  components of the model, highlighting the critical aspects for aeroacoustic applications. 
  The paper is organized as follows. We start with the overview of the numerical method in section II where we will describe all the
  different aspects of the model and its optimization for aeroacoustic computations. We show test results of our model on the two study cases
  for both the flow field and the acoustic field in section III and IV. Finally, in section V we draw our conclusions.

\section{THE NUMERICAL METHOD}

\subsection{Standard and optimized Lattice Boltzmann dynamics}
  
  The most widely Lattice Boltzmann methodology, known as an efficent solver for weakly compressible isothermal fluid flows,
  is based on a fully discrete version of the BGK Boltzmann equation (e.g. Chen and Doolen\citep{Chen1998}). A projection onto a finite number
  of discrete velocities directions, the lattice, and a discretization of the advection term gives the classical
  Lattice Boltzmann BGK evolution equation (LBM-BGK): 
  \begin{equation}
    \label{standardLB}
    f_i(\bm{x}+\bm{\xi_{i}}\delta{t},t+\delta{t}) = f_i(\bm{x},t) -\frac{1}{\tau}(f_i(\bm{x},t)-f^{eq}_i(\bm{x},t)) 
  \end{equation}
  where  $i = 1\ldots q$ are the different lattice directions, $f_i$ is the discrete probability density to find a particle at position $x$ and time $t$
  with velocity $\bm{\xi_{i}}$ (along the direction i), $f^{eq}_i$ is the discrete equilibrium distribution function, $\tau$ the relaxation time
  and $\delta{t}$ the time step. The macroscopic quantities, like the fluid density $\rho$ and velocity  $\bm{u}$, can be obtained via the
  discrete moments of the distribution functions
  \begin{equation}
   \label{moments}
    \rho=\sum_{i}f_{i}(\bm{r},t) \hspace{2em} \bm{u} = \frac{1}{\rho} \sum_{i}f_{i}(\bm{r},t)\bm{\xi_{i}}
  \end{equation}
  The $f^{eq}_i$ in the LBM-BGK is commonly represented by a quadratic polynomial approximation of the continuous Maxwell-Boltzmann
  equation:
  \begin{equation}
    \label{truncatedEq}
    f^{eq}_i = w_i\rho(1+\frac{\bm{u}\cdotp\bm{\xi_i}}{c_{s}^2}+\frac{(\bm{u}\cdotp\bm{\xi_i})^{2}}{2c_{s}^4}
    -\frac{|\bm{u}|^2}{2c_{s}^2})+O(\bm{u}^3)
  \end{equation}
  where the weights $w_i$ and the lattice sound speed $c_s$ are coefficients specific to the lattice topology. 
  This approximation provides the possibility to numerically solve eq. \ref{standardLB} but also poses some important limitations
  for the LBM-BGK to resolve the fluid dynamics at the macroscopic level. 
  First, the model is valid only for isothermal computations with the pressure $p=c^2_{s}\rho$ and the bulk viscosity $\nu'$ is equal to $2/3$ of
  the shear kinematic viscosity $\nu$.\citep{Dellar2001} In particular, the shear viscosity is directly linked to the relaxation
  time trough $\nu=c^{2}_{s}(\tau-1/2)\delta{t}$.
  Second, the stress ($\bm{\sigma}$) - strain rate ($\bm{S}$) relationship of the retrieved macroscopic equations deviates
  from the one for Newtonian fluids due to the presence of an additional spurious term:\cite{Qian1998}
  \begin{equation}
    \label{constitutiveEq}
    \bm{\sigma} = \rho\nu[(\nabla\bm{u})+(\nabla\bm{u})^{T}]-\tau\nabla\cdotp(\rho\bm{u}\bm{u}\bm{u})
  \end{equation}
  The term $\nabla\cdotp(\rho\bm{u}\bm{u}\bm{u})$ is in fact an error terms that breaks the Galilean invariance of the retrieved
  macroscopic equations \cite{Qian1998}.
  As it scales as $\mathrm{Ma}^3$ (the error as $\mathrm{Ma}^2$) has been considered negligible for simulations with sufficiently
  small Mach numbers. However, it can strongly affect the accuracy and the numerical stability of the LBM-BGK at high Reynolds Number.
  \citep{Ricot2009}${}^,$\citep{Dellar2014}${}^,$\citep{Malaspinas2015} As already explained above,
  this limitation become even more restrictive for aeroacoustic simulations because the appearance of numerical instability waves
  can deteriorate significantly the quality of acoustic predictions, even at low $\mathrm{Ma}$. In order to alleviate this problem,
  Malspinas\citep{Malaspinas2015} has recently proposed
  a smart optimization of the LBM-BGK. Here we review the basic idea.
  
   We start considering the continuous distribution function $f(\bm{x},\bm{\xi},t)$ which can be expanded in Hermite polynomials up
 to a order $N$:\citep{Shan2006}${}^,$\citep{Grad1949}${}^,$\citep{Grad1949b}
 \begin{equation}
 \label{HE}
   f(\bm{x},\bm{\xi},t) \approx f^{N}(\bm{x},\bm{\xi},t) = \omega(\bm{\xi})\sum_{n=0}^{N}\frac{1}{n!} \mathcal{H}^{n}(\bm{\xi}):\bm{a}^{n} 
 \end{equation}
 where the colon symbol '$:$' stands for the full index contraction, $\mathcal{H}^{n}$ for the Hermite polynomials of order
 n and $\omega = \exp(-\bm{\xi}^{2}/2)$ for the associated Gaussian weight.
 For a complete review on the Hermite polynomials the interested reader is referred to Shan \textit{et al.}\cite{Shan2006}
 The expansion coefficients $\bm{a}^{n}$ are defined by 
 \begin{equation}
 \label{Hcoeff}
  \bm{a}^{n}(\bm{x},t) = \int f(\bm{x},\bm{\xi},t)\mathcal{H}^{n}(\bm{\xi})d{\bm{\xi}}
 \end{equation}
 and correspond exactly to the velocity moments of the degree $n$. As a result, the observable macroscopic
 variables are directly obtained by the first three coefficients (four considering thermal fluids):
 \begin{equation}
  \bm{a}^{0} = \rho;  \  
  \bm{a}^{1} = \rho\bm{u}; \
  \bm{a}^{2} = \bm{P}-\rho(\bm{uu}-\delta);
 \end{equation}
 where   $\bm{\sigma} = \bm{P}-\rho(\bm{uu}-\delta)$ is the stress tensor.
 Due to the orthogonality of the Hermite polynomials, the velocity moments of $f$ and $f^N$ are the exactly the same
 (up to the order n). As an example, a truncated Hermite expansion of $f$ up 3rd order $f^{3}$ allows to
 approximate the distribution function retaining the information necessary for an exact evaluation of its first three moments.

 Before proceeding with the critical step of the discretization in the velocity space, we expand the distribution
 function as a perturbation of the equilibrium distribution (Chapman-Enskog expansion):
 \begin{equation}
 \label{CE}
  f = f^{eq}+f^{neq}
 \end{equation}
  where the suffix $neq$ stand for the non equilibrium part.
  Naturally,  $f^{eq}$ and $f^{neq}$ can be also expressed in terms of Hermite polynomials:
 \begin{equation}
  \label{HEeqAndNeq}
  f^{eq,N}=w_{i}\sum_{n=0}^N \frac{1}{n!} \mathcal{H}^{(n)}(\bm{\xi}):\bm{a}^{n}_{eq}
    \hspace{2em} f^{neq,N}=w_{i}\sum_{n=0}^N \frac{1}{n!} \mathcal{H}^{(n)}(\bm{\xi}):\bm{a}^{n}_{neq}
  \end{equation}
  where $\bm{a}^{n}_{eq}$ and $\bm{a}^{n}_{neq}$ are defined by eq. \ref{Hcoeff} with $f^{eq}$ and $f^{neq}$ in place of $f$. 
  Now, it is important to note that there exist recursive formulations for both the equilibrium and non-equilibrium
  expansion coefficients:\citep{Malaspinas2015}
  \begin{equation}
  \label{recursive1}
  \bm{a}^{n}_{eq} = \bm{a}^{n-1}_{eq}\bm{u},\qquad  \bm{a}^{0}_{eq} = \rho; \hspace{2em}
  \end{equation}
  and 
  \begin{equation}
    \label{recursive2}
    a^{n}_{neq,\alpha_{1}...\alpha_{n}} = a^{n-1}_{neq,\alpha_{1}...\alpha_{n-1}}u_{\alpha_{n}}
    +(u_{\alpha_{1}...\alpha_{n-2}} a^{2}_{neq,\alpha_{n-1}\alpha_{n}} +perm(\alpha_{n}) ).
  \end{equation}  
  where “$perm(\alpha_{n})$” stands for all the cyclic index permutations of indexes from
  $\alpha_{1}$ to $\alpha_{n-1}$. In addition we know that $a^{0}_{neq}$ and  $a^{1}_{neq}$ are null in order
  for the macroscopic equations to satisfy mass and momentum conservation, respectively.
  Equations (\ref{recursive1} and \ref{recursive2}) interestingly allow the reconstruction of $f^{eq}$ and $f^{neq}$ up
  to any order $N$ in the Hermite expansion, just knowing $\rho$, $\bm{u}$ and  $\bm{a}^{(2)}_{neq}$. 
  As it will be more clear later on, these recursive relationships are the key 'ingredient' of the new model.
  
  We can now discuss the discretization of the velocity space.
  As already stated above, a truncated Hermite expansion of $f$ up $N$-th order $f^{N}$ allows us to approximate the distribution
  function retaining the information necessary for an accurate evaluation of its first N-moments, ( i.e. the macroscopic
  fluid variables we are interested in). In turn, a truncated Hermite expansion $f^N$ of $f$,
  can be completely defined, via a Gauss–Hermite quadrature which substitute $f^N(\bm{\xi})$ with
  function values on a set of discrete microscopic velocities $f^{N}(\xi_i)$. If a quadrature formula of sufficient
  precision is used, the evolution equation of the discrete set $f^N_{i} \equiv f^N(\bm{\xi}_{i})$ provides the exact evolution 
  of the first $N$ moments of $f$ and therefore of the corresponding macroscopic variables.
  The most commonly used quadratures, which correspond to the lattices D2Q9 (in two dimensions) and the D3Q15,
  D3Q19, and D3Q27 lattices\cite{DdQq}
  (in three dimensions), allow the exact representation of $f$ only up to the second order in Hermite polynomials. 
  However, Shan \textit{et al.}\cite{Shan2006} has demonstrated that the continuous Maxwell–Boltzmann equilibrium distribution $f^{eq}$ must be
  approximated in Hermite expansions up to the third order to accurately recover the isothermal Navier–Stokes dynamics.
  Thanks to the recursive formulations for the Hermite coefficients and a special property 
  of the D2Q9 and D3Q27 lattices, Malaspinas\cite{Malaspinas2015} has proposed to expand the
  discrete equilibrium and the non equilibrium distribution to order $n > 2$ in Hermite polynomials:
   \begin{equation}
   \label{eqH}
   f^{eq\mathcal{H}}_i=w_{i}\rho(\underbrace{1+\frac{\bm{\xi}_{i}}{c_s^2}+\frac{1}{2c_s^4}\mathcal{H}^{(2)}_i:\bm{u}\bm{u}}_{\mbox{standard part}}
   +\underbrace{O(\mathcal{H}^{(3)}_i:\bm{u}\bm{u}\bm{u})+\dots}_{\mbox{extended equilibrium} }).
   \end{equation}
    and 
   \begin{equation}
   \label{neqH}
   f^{neq\mathcal{H}}_i=w_{i}\rho(\underbrace{\frac{1}{2c_s^4}\mathcal{H}^{(2)}_i:\bm{a}^{2}_{neq,i}}_{\mbox{standard part}}
   +\underbrace{O(\mathcal{H}^{(3)}_i:\bm{a}^{3}_{neq,i})+\dots}_{\mbox{extended part} }).
   \end{equation}
   where the Hermite coefficients of the equilibrium distribution are the ones obtained from of the continuous Maxwell-Boltzmann distribution
   for both the 2D and 3D cases, simplifying the computations. Instead, the coefficients for $f^{neq\mathcal{H}}_i$ are determined
   using their recursive relationship. The complete expression of eq. \ref{recursive1} and \ref{recursive2} 
   can be found in Malaspinas.\cite{Malaspinas2015}
   Nonetheless, since we now expand the distribution function up to a limited order
   in Hermite polynomials the recursive relations are not anymore exactly verified.   
   With the discretization (the quadrature), eq. \ref{recursive1} and \ref{recursive2} are not the same but present some
   error terms $O(\mathrm{Ma}^{n+1})$, where $\mathrm{Ma}$ is the Mach Number. These spurious terms are assumed to be small since they are one order higher in physical velocity.
   The detailed expressions of the recursive relations for the discrete case can  also be found in 
   Malaspinas.\cite{Malaspinas2015}
   Finally, an evolution equation for the new model is written as     
  \begin{equation}
    \label{optimizedLB}
    f_i(\bm{x}+\bm{\xi_{i}}\delta{t},t+\delta{t}) = f^{eq\mathcal{H}}_i(\bm{x},t) + (1-\frac{1}{\tau})f^{neq\mathcal{H}}_i(\bm{x},t)
  \end{equation}
  
  A Chapman-Enskog expansion of this model reveal that the effect of the erroneous cubic term on
  the stress tensor (\ref{constitutiveEq}) $\bm{\sigma}$ is greatly reduced. In particular, the dispersion on the 
  diagonal components of $\bm{\sigma}$ is diminished and the off-diagonal components are completely correct. The overall effect is to 
  increase not only the accuracy but also the stability of the standard LBM. This fact is confirmed also by the good dispersion and
  dissipation properties shown by the optimized model when analyzed with Von Neumann linear
  stability analysis.\citep{Malaspinas2015}
  Also when tested on benchmark problem the LBM-rrBGK presents increased accuracy and stability over the LBM-BGK and LBM-MRT.
  
  In practice, eq. \ref{optimizedLB} will be resolved following the regularized procedure of Latt and Chopard:\cite{Latt2006}   
  during a simulation at each iteration, before a new collision step is performed, the equilibrium 
  and non equilibrium part of the distribution function are regularized recalculating them through their Hermite formulation 
  (eq. \ref{eqH} and \ref{neqH}) with the current values of $\rho$ and $\bm{u}$. In this way, the numerical scheme will be
  forced to be consistent with a theoretical framework which more accurately recover the stress tensor $\bm{\sigma}$.
 
\subsection{Multi-Domain Grid Refinement}
  
  In a multi-domain approach refined grids patches occupy almost distinct areas of the computational domain
  and are assembled like in a jigsaw puzzle. 
  As in most LBM, a regular grid is used and a drastic change in scale happens at the grid transition.
  Since each resolution level possesses its own units, the simulation variables have to be rescaled before
  communicating them between grids at different resolution.
  Following Lagrava \textit{et al.},\cite{Lagrava2012} we choose $\delta{x_f} = \delta{x_c}/2$ to be the spatial
  discretization of the fine grid respect to the coarse grid. 
  From here on we will refer in general to coarse grid units with the $c$ subscript and with the $f$ subscript to
  fine grid units.
  In order to define the temporal scale we use a convective scaling $\delta{x}\sim\delta{t}$. Given these conditions,
  the ratio of the spatial and the temporal scale is constant and subsequently one time iteration on the coarse grid
  corresponds to two iterations on the fine grid:
  \begin{equation}
    \label{convective}
    \frac{\delta{t_f}}{\delta{x_f}} = \frac{\delta{t_c}}{\delta{x_c}} = const. \hspace{1em} \Longrightarrow \hspace{1em} \delta{t}_f = \delta{t}_c/2
  \end{equation}
  In addition, the convective scaling implies also that the velocity, the density and the pressure fields are continuous
  on the grid transition. In other words, these physical quantities do not need any rescaling and therefore:
  \begin{equation}
  \label{macroscopicContinuity}
  \rho = \rho_c =\rho_f \hspace{1em} \hspace{1em}  \bm{u} = \bm{u_c} = \bm{u_f} 
  \end{equation}
   The viscosity, instead, depends on the grid resolution and has to be rescaled according to
   $ \nu_f=\frac{\delta{x_c}}{\delta{x_f}}\nu_c$. As a result, the relaxation time 
   follows the rescaling $\tau_f = (4\tau_c -1)/2$.\citep{Lagrava2012}
  For the rescaling of the distribution function we adopt the method proposed for the first time by 
  Filippova and H{\"a}nel,\cite{Filippova1998} later modified by Dupuis and Chopard.\cite{Dupuis2003}
  The modification proposed by  Dupuis and Chopard\cite{Dupuis2003} consists in rescaling $f_i$ before
  the collision step in order to avoid limitations on the value of the relaxation time. As noted before,  
  the populations $f_i$ can be expressed as the sum of an equilibrium and non-equilibrium part.
  In case, we adopt the LBM-rrBGK:   
  \begin{equation}
    f_{i,n} = f^{eq\mathcal{H}}_{i}(\rho,\bm{u})+f^{neq\mathcal{H}}_{i,n}(\bm{\nabla}\bm{u})_n
  \end{equation}
  where the subscript $n$ can stand for $c$ (coarse) or $f$ (fine). The equilibrium distribution $f^{eq\mathcal{H}}_i$,
  is a function
  of $\rho$ and $\bm{u}$ only and can be evaluated independently of the grid resolution ($f^{eq}_{i,n}=f^{eq}_i$,
  see eq. \ref{macroscopicContinuity}).
  The non equilibrium part instead is a function of the gradient of $\bm{u}$ whose value is grid dependent. As a result,
  at the grid transition only $f^{neq\mathcal{H}}_{i}$ needs to be rescaled. Considering the Chapman-Enskog 
  leading order approximation
  of the $f^{neq\mathcal{H}}_i$,\citep{Dupuis2003} the rescaling of $f^{neq\mathcal{H}}_{i}$ can be written as:
  \begin{equation}
    f^{neq\mathcal{H}}_{i,f} = \alpha f^{neq\mathcal{H}}_{i,c}  
  \end{equation}
   where $\alpha = 2\frac{\tau_f}{\tau_c}$ is the scaling factor for a temporal refinement $\delta{t_f} = \delta{t_c/2}$
   (eq. \ref{convective}).  
   Once the rescaling of the simulation variables is correctly defined, one still requires an 
   algorithm for the grids coupling, which specifies the procedure to pass the rescaled variables from one resolution level
   to the other. The two way coupling proposed by Lagrava \textit{et al.}\cite{Lagrava2012} requires a small layer of overlap between neighboring
   grids as drawn in fig. 1.
   \begin{figure} % 1
        \centering
         \includegraphics[width=0.6\textwidth]{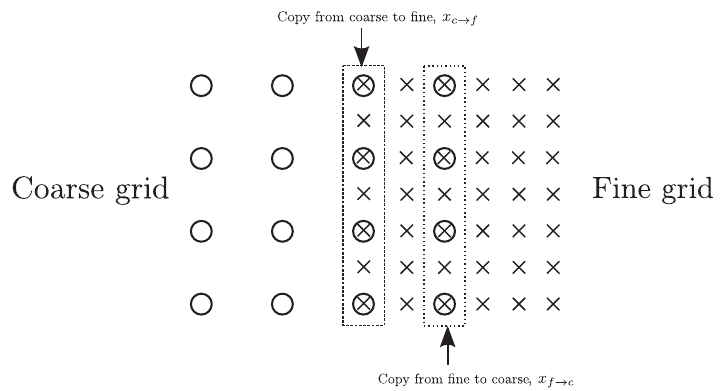}
          \caption*{Figure 1. Sketch of the overlap region between coarse and fine grids from Lagrava \textit{et al.}.\citep{Lagrava2012}
            The dotted line highlights the region where the information is exchanged.}
%                  \label{gridsOverlap}
   \end{figure}   
   From this figure, it is clear that when passing from the fine to the coarse
   grid the amount of information has to be reduced by decimation, while it must be increased, or in some way reconstructed,
   when going from the coarse to the fine grid. Therefore, a two way coupling involves a decimation and a reconstruction operation.

   For the decimation, a low pass filtering has been shown to be an optimal choice when dealing with turbulent flows,
   especially at high Reynolds number, since completely removes the effect of the small scales that cannot be resolved
   by the coarse grid. The possible presence of these small scales in the coarse grid often acts as a trigger
   of numerical instabilities. 
   It is also worth mentioning that the filtering operation has to be applied only on $f^{neq\mathcal{H}}_i$.
   A direct filtering of the complete $f_i$ or just of $\rho$ and $\bm{u}$ has been shown to introduce an artificial dissipation
   which causes loss of accuracy and an unphysical increase of the viscosity.\citep{Lagrava2012}
   
   The reconstruction is more complex. This operation requires to reform the missing populations $f_i$
   and macroscopic variables $\rho$ and $\bm{u}$ for the sites on the fine grid $\bm{x}_f$ not having any corresponding sites
   on the coarse grid $\bm{x}_c$ (fig. 1). The missing information for these sites can be reconstructed 
   by an interpolation of the simulation variables in the neighboring coarse sites, $\widetilde{f_{i,c}}$, $\widetilde{\rho_c}$
   and $\widetilde{\bm{u_c}}$. Regarding accuracy, the spatial interpolation is the most critical part of the
   grid refinement algorithm. In particular, numerical investigations have shown that at least a local cubic interpolation
   is necessary to retain the second order accuracy of the LBM.\citep{Lagrava2012} A lower order interpolation typically
   gives rise to artificial sharp discontinuities at the grid interfaces,  which affect not only the accuracy but also
   the stability. 
   
   Finally, the above considerations for the two way coupling can be summarized into two equations that allow us to transform
   the fine and coarse distributions into their corresponding partners in the overlap region  (fig. 1):
   \begin{equation}
     \label{couplingCF}
     f_{i,c}(\bm{x}_{f\rightarrow c}) = f^{eq\mathcal{H}}_i(\overline{\rho}(\bm{x}_{f\rightarrow c}),\overline{\bm{u}}(\bm{x}_{f\rightarrow c}))
     +\overline{f^{neq\mathcal{H}}_{i,c}}(\bm{x}_{f\rightarrow c})
   \end{equation}
   
   \begin{equation}
      \label{couplingFC}
     f_{i,f}(\bm{x}_{c\rightarrow f}) = f^{eq\mathcal{H}}_i(\widetilde{\rho}(\bm{x}_{c\rightarrow f}),\widetilde{\bm{u}}(\bm{x}_{c\rightarrow f}))
     +\widetilde{f^{neq\mathcal{H}}_{i,c}}(\bm{x}_{c\rightarrow f})
   \end{equation}   
   For a correct interpretation of the index notation the reader is referred to fig. 1.
   The over-line (e.g. $\overline{f^{neq\mathcal{H}}_i}$) stands for low-pass filtering and the tilde
   (e.g. $\widetilde{f^{neq\mathcal{H}}_i}$) for interpolation.  
   Regarding the coupling for the sites $\bm{x}_{f\rightarrow c}$ (eq. \ref{couplingCF}) there are two different cases.
   If a fine site corresponds to a coarse site, the coupling consists in a simple rescaling of the non-equilibrium distribution
   $\widetilde{f^{neq\mathcal{H}}_{i,c}}=\alpha f^{neq\mathcal{H}}_{i,f}$
   and copying the $\rho_c$ and $\bm{u}_c$ to obtain $f^{eq}$. When instead, a fine site does not correspond to any coarse site
   $\bm{x}_{f\rightarrow c}$ the coupling requires the calculation of the interpolated variables
   $\widetilde{f^{neq\mathcal{H}}_{i,c}}$, $\widetilde{\rho}$ and $\widetilde{\bm{u}}$. 
   
   Eq. \ref{couplingCF} and \ref{couplingFC} together ensure the continuity at the grid transition for both the mesoscopic and macroscopic fields.
   However, this continuity does not hold for any time step. When a collide-stream operation is executed, the distributions
   of the coarse grid and the fine grid are updated at two different time steps (see eq. \ref{convective} ), $t+\delta{t}_c$ and
   $t+(\delta{t}_{c}/2)$ respectively. The fine sites at the time $t+(\delta{t}_{c}/2)$ miss the information coming from the
   coarse sites. In order to reconstruct the absent populations $f_{i,c}(t+\delta{t}_{c}/2)$ one can apply a linear
   interpolation in time (which is second order accurate at $\delta{t}_{c}/2$) on the coarse sites of the overlap region (fig. 1).
   With the time interpolation both the coarse and the fine grids are complete and ready for a new collide and stream (iteration).
   
   The presented grid refinement algorithm has been tested successfully with the standard LBM on a classical acoustic benchmark
   (propagation of an acoustic pulse) by Gendre \textit{et al.}\cite{Gendre2017} Using a different benchmark,
   the same authors have recently highlighted that when a vortical structure cross the refinement interface may produce
   spurious acoustic waves contaminating the wave field of interest. This effect can be reduced using a directional
   splitting approach at the refinement interface\citep{Gendre2017} and will be considered in future
   optimizations of the present model.

\subsection{Non reflective boundary conditions: an optimized absorbing boundary layer for LBM.}
  
   Regardless of the numerical method, the disturbances approaching the computational boundaries can be suppressed by adding
   a damping term to the governing equations. The role of this damping force is to drive
   the computed solution towards a prescribed stationary solution. In its most popular form, the force term is proportional to the
   difference between the actual value and a pre-defined reference value of any physical variable $q$  (e.g pressure):
   \begin{equation}
   \label{dampingForce}
    -\chi(x)(q-q_{ref})
   \end{equation}
   where the damping intensity $\chi(x)$, the so called 'sponge strength', ranges from zero in the physical domain to its maximum value %(any positive number)
   in the sponge layer.  The classical absorbing theory predicts an exponential decay for the amplitude of the disturbances passing trough
   the sponge layer, if the value of $\chi$ is sufficiently large.\citep{Israeli1981}${}^,$\citep{Wagner2007}${}^,$\citep{Freund2000}
   
   A rigorous definition of the sponge layer for LBM has been attempted by Xu and Sagaut.\cite{Xu2013} Three different possible formulations have been
   explored and compared but, on the base of theoretical and numerical investigations, only one turns out to be more/unconditionally stable.
   Here we describe the basic idea only for this optimal absorbing strategy adapted to the optimized dynamics.
   
   As for the standard LBM-BGK (\ref{standardLB}), also the LBM-rrBGK (\ref{optimizedLB}) can be rewritten
   with the additional damping term $F_i$: 
   \begin{equation}
   \label{spongeLB}
   f_i(\bm{x}+\bm{\xi_{i}}\delta{t},t+\delta{t}) = f^{eq\mathcal{H}}_i(\bm{x},t) + (1-\frac{1}{\tau})f^{neq\mathcal{H}}_i(\bm{x},t)
   +\delta{t}F_{i}.
   \end{equation}
   The general force term $F_i$, in analogy with eq. \ref{dampingForce}, can be defined by:
   \begin{equation}
     F_i = \chi(f^{ref}_{i}-f^{*}_{i})
   \end{equation}
   where $f^{ref}_{i}$ is the reference state for $f_i$ (e.g time average value of $f_i$) and $f^{*}_i$ the possible representation of the
   mesoscopic variables. One can choose to express both $f^{ref}_i$ by the equilibrium distribution, which directly depends
   on $\rho^{ref}$ and $\bm{u^{ref}}$. 
   In this way, instead of dealing with the mesoscopic quantities we only need to handle macroscopic variables
   (the reference fields), which is simpler and more effective. In fact, in aeroacoustic applications the reference
   fields are often taken as the far field or the mean fields of the macroscopic variables that can be used directly
   in the LBM if $f^{ref} \equiv f^{eq}(\rho^{ref},\bm{u}^{ref})$.
   Likewise,   $f^{*}_i$ has to be defined by the equilibrium distribution function 
   to avoid possible numerical instabilities.
   Finally, also from the computational point of view this choice is convenient for saving both memory and CPU cost.
   
   Based on these considerations the force term can be defined as:
   \begin{equation}
   \label{dampingLBM}
    F_i = \chi(f^{eq\mathcal{H}}_{i}(\rho^{ref},\bm{u}^{ref})-f^{eq\mathcal{H}}_{i}(\rho^{*},\bm{u}^{*})).
   \end{equation}
   where $\rho^{ref}$ and $\bm{u}^{ref}$ are the reference values for the density $\rho^{*}$ and the velocity fields $\bm{u}^{*}$ that,
   according to the dynamics of eq. \ref{spongeLB}, in the sponge layer ($\chi \neq 0$) have to be computed with the following
   equations:\cite{Guo2002}   
  \begin{center}
  \label{modifiedMoments}
    $\rho^{*}=\sum_{i}f_{i}(\bm{r},t)+n\delta{t}\sum_{i}F_i$ \hspace{1em} 
    $\bm{u}^{*} = \frac{1}{\rho*} \sum_{i}f_{i}(\bm{r},t)\bm{\xi_{i}}+m\delta{t}\sum_{i}F_i\bm{\xi_{i}}$
  \end{center}
  In the physical domain $\chi = 0$, the dynamics expressed by eq. \ref{spongeLB} become equals to the LBM-rrBGK eq.
  \ref{optimizedLB} and the macroscopic variables are retrieved with the original form (eq. \ref{moments}).
  
  A Chapman-Enskog expansion of eq. \ref{spongeLB} with the damping term (eq. \ref{dampingLBM}) provides the
  corresponding macroscopic equations in the sponge layer. For the purposes
  of the present discussion, we report the macroscopic equations in a simplified (1D linearized) form 
  for the classical LBM-BGK:\cite{Xu2013} 
  \begin{equation}
    \begin{sistema}
    \partial_{t}\rho'_{x}+\partial_{x}u'_{x} = -\sigma^{eff}\rho' \vspace{1em} \\ 
    \partial_{t}u'_{x}+{c^{eff}_{s}}^{2}\partial_{x}\rho' = -\sigma^{eff}\rho'
    \end{sistema}
  \end{equation}
  where $\rho'= \rho^{*}-\rho^{0}$ and $u'= u^{*}-u^{0}$ are small fluctuation on the mean fields $\rho^{0}= 1$ and $u^{0}=0$
  (e.g. in the far field). It is important to say that the effective damping coefficient $\sigma^{eff}$
  and the effective sound speed $c^{eff}_{s}$ are coupled with both the sponge strength $\chi$ and relaxation time $\tau$:
   \begin{equation}
   \label{effectiveCoefficient}
    \sigma^{eff} = \frac{(1+1/2\tau)\chi}{(1-\chi/4\tau)}  \hspace{3em}
    {c^{eff}_{s}}^2 = \frac{c^{2}_{s}(1-\chi\tau)}{(1-\chi/4\tau)}
   \end{equation}.
  The correct effective sound speed is obtained only when $\tau=1/2$. However, there is no requirement
  for the computed solution to be physical in the sponge layer. Therefore also $\tau\neq1/2$ can be possible.
  This statement remains true as long as the unphysical sponge layer does not introduce
  significant reflections in the physical domain.
  
  Based on eq. \ref{effectiveCoefficient} we can determine a critical value of $\chi$ for the stability of the absorbing scheme
  and optimize the use of the sponge layer for LBM.   
  In fact, for stability $\sigma^{eff}$ has to be greater than zero and therefore $\chi < 4\tau$. 
  which provides an upper limit when choosing a reasonable value for the absorbing strength $\chi$.
  The above theoretical considerations on the stability are further confirmed when the dissipative and dispersive
  properties are explored through the Von-Neumann stability analysis.\citep{Xu2013}
  It is worth noting that in the absorbing theory for the Navier-Stokes equations, $\sigma$ is a penalized
  parameter, which in theory can assume any real positive value. In practice it is not chosen very large.
  In the absorbing theory for LBM instead $\sigma$ is a finite parameter coupled with the relaxation 
  time.\citep{Xu2013}
  
  Regardless of the method used to solve the fluid motion, an uniform profile for the sponge profile $\sigma(\bm{x})$
  can cause significant reflections. 
  As an example, if $\sigma(\bm{x}) = const.$ even a low damping target\cite{DampingTarget} (say 20 dB) 
  is achivied when the sponge layer lenght $L_{sp}$ is more than 10 times the incident wave lenght $\lambda$.\citep{Mani2010}
  This condition renders the sponge layer practically ineffective but can be significantly relaxed to $L_{sp}\approx 2*\lambda$
  if non uniform absorbing profiles $\sigma(\bm{x})$ are adopted.
  In this work we adopted a polynomial profile for $\sigma(\bm{x})$:\citep{Xu2013}${}^,$\citep{Israeli1981}
  
     \begin{equation}
      \sigma(\bm{x})=\frac{3125(L-x)(x-x_{0})^4}{256(L-x_{0})^5}
     \end{equation}

 \section{BENCHMARK: SOUND GENERATED BY AN INFINITE CYLINDER}
 
    Our purpose in this section is to test our model capabilities in reproducing the generation and
    the propagation of the sound by a two-dimensional circular cylinder in an uniform flow. This problem has been
    studied extensively by many authors through theoretical, experimental and numerical investigations 
    (e.g.  Strouhal,\cite{Strouhal1878} Curle,\cite{Curle1955} Williamson,\cite{Williamson1996}
    Colonius \textit{et al.}\cite{Colonius1994}).
    In this work we will refer mainly to the DNS study of Inoue and Hatakeyama,\cite{Inoue2002} where the flow and
    acoustic fields are computed directly including both the near and the far fields in the computational domain.
  
    The Mach Number ($\mathrm{Ma}$) and the Reynolds number ($\mathrm{Re}$) of the problem are defined by $Ma = U_{\infty}/c_{\infty}$ and  
    $\mathrm{Re} = U_{\infty}D/\nu_{\infty}$, where $U_{\infty}$ is the velocity of the uniform flow, $D$ the cylinder diameter,
    $c_{\infty}$ the ambient sound speed (equal to $c_s$ in LBM units) and $\nu_{\infty}$ the kinematic viscosity.
    Following Inoue and Hatakeyama,\cite{Inoue2002} we have chosen 
    $\mathrm{Ma} = 0.2$ and $\mathrm{Re} = 150$. The value of the Reynolds number is sufficiently high to allow the formation of the vortex shedding
    responsible for the sound generation but still small enough to avoid the transition to a turbulent flow ($\mathrm{Re} >160$).   
    The computational setup is reported in fig. 2.
    \begin{figure}  % 2
        \centering
         \includegraphics[width=0.5\textwidth]{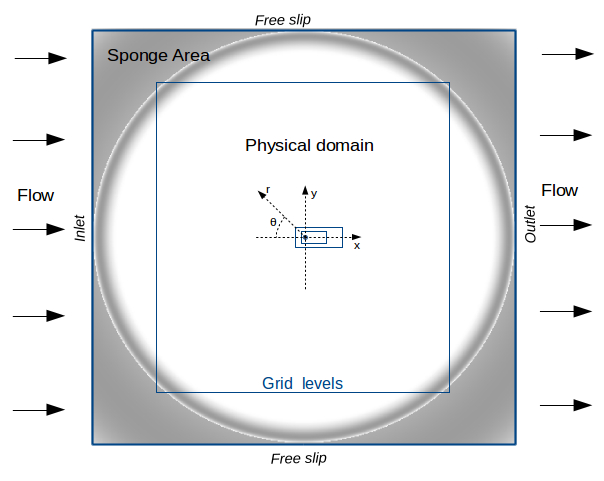}
         \caption*{Figure 2. Simulation setup for the sound generated by a uniform flow around a two dimensional cylinder. A gray scale is used to visualize the sponge strenght. 
          The latter is zero in the physical domain (white) and gradually increases near the domain boundaries (dark gray).
          }
%        \label{cylinderSetup}
   \end{figure}
    The domain size is $[300\times300]\times D$ and divided
    into $4$ refinement levels. The finest resolution level ($\delta{x} = D/N = 0.03125 $ and $\delta{t} = 0.00361$,
    with $D=1$ the adimensional diameter and N the number of points) is located near the
    circular walls of the cylinder. 
    The thickness of the boundary layer is  approximately $\delta/D \sim 0.08$ and in our simulation remains under-resolved.
    The coarsest refinement level ($\delta{x} = 0.25$, $\delta{t} = 0.02887$) is placed in the outer part of the computational domain and include a 
    sponge layer with a minimum width $L_{sp} = 20D$. Preliminary tests have also shown this value to be enough for the sponge layer
    to damp the acoustic disturbances reaching the computational boundaries, thus preventing spurious reflections. Although not reported here,
    smaller values for $L_{sp}$ are also possible.  The sponge layer is circular. In this way,
    the angle of incidence of the wave field with respect to the sponge layer is always small,
    which is optimal to maximize the damping efficiency.\citep{Mani2010} In addition, the corners of the computational
    domain are covered by the sponge layer to avoid the appearance of spurious low frequency waves.  
    Regarding the boundary conditions, we used free slip condition (zero-gradient for tangential velocity components) for the
    lateral boundaries,  Dirichlet condition for the inflow (left boundary) and outflow condition (zero-gradient for all
    velocity components) for the right boundary. The flow field is initialized as an uniform flow field with velocity
    $U_{\infty}$ and no perturbation is added to achieve a faster transition to the vortex shedding regime.
  
    For this problem, the fluctuation of the lift force, related to the vortex shedding, is much bigger than the drag fluctuation and
    is mainly responsible for the generated sound field. In our simulation, the peak amplitude of the lift coefficient
    ($C_{L}\sim 0.55$ for N=32)
    is close to both numerical ($0.52$ for $\mathrm{Re} = 150$\citep{Inoue2002}) and experimenta values
    ($0.48-0.55$, for $\mathrm{Re} = 140-160$\citep{Kwon1996}). 
    The Strouhal number ($ \mathrm{St} = f D/ U_{\infty} $, where f is the frequency) for the oscillation of $C_L$ is $0.183$
    and is in agreement with the values found in the literature.\citep{Inoue2002}${}^,$\citep{Williamson1993}
    Results for the flow field are visualized by contours of vorticity ($\omega$), while  the acoustic field is
    represented by contours of fluctuating pressure $\Delta{\tilde{p}}$ (fig. 3), as
    defined by Inoue and Hatakeyama:\cite{Inoue2002}
    \begin{equation}
    \Delta{\tilde{p}} = \Delta{p} - \Delta{p}_{mean}
    \end{equation}
    where $\Delta{p} = p-p_{amb}$, with $p_{amb}$ the ambient pressure and $\Delta{p}_{mean}$ the temporal mean of $\Delta{p}$.
    \begin{figure}  % 3
         \centering
         \includegraphics[height=0.4\linewidth]{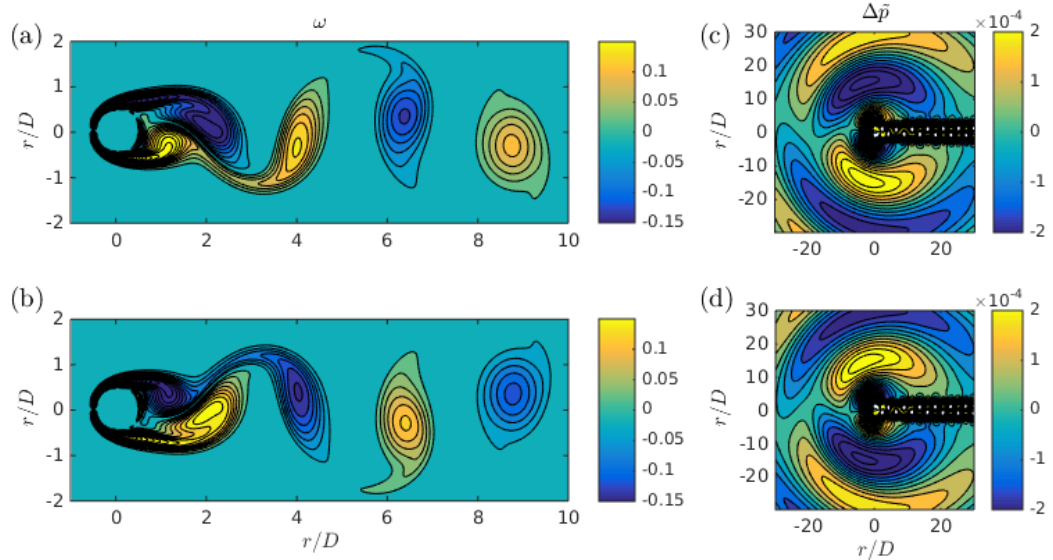}
         \caption*{Figure 3.(Color online) On the left the time development of the vorticity field $\omega$ for two successive time step: (a) $t=t_1$, (b) $t=t_{1}+\Delta{t}$ with $\Delta{t}=2.7$) .
         On the right the corresponding time development of the fluctuation pressure field $\Delta{\tilde{p}}$: (c) $t=t_1$, (d) $t=t_2$.
         The contour levels are from $\omega_{min}=-1.0$ to $\omega_{max}=1.0$ with an increment of 0.02 and from $\Delta{\tilde{p}}_{min}=-0.1{\mathrm{Ma}}^{2.5}$
          to $\Delta{\tilde{p}}_{min}=0.1{\mathrm{Ma}}^{2.5}$ with an increment of $0.0025{\mathrm{Ma}}^{2.5}$.}
%         \label{cylinderNearField}
    \end{figure}
    The well known sound source mechanism appears to be correctly reproduced (fig. 3):
    an alternate vortex shedding from the upper and bottom sides of the cylinder produces oscillating pressure fluctuations
    (with opposite phases) on the lower and the upper sides of the cylinder, which in turn propagate as pressure waves away
    from the solid body. The resulting wave field is reported in fig. 4a, where we compare the isocountours 
    of $\Delta{\tilde{p}}$ with the ones obtained by the DNS investigation\cite{Inoue2002} (fig. 4b).
    \begin{figure}  % 4
      \centering
       \includegraphics[width=0.7\linewidth]{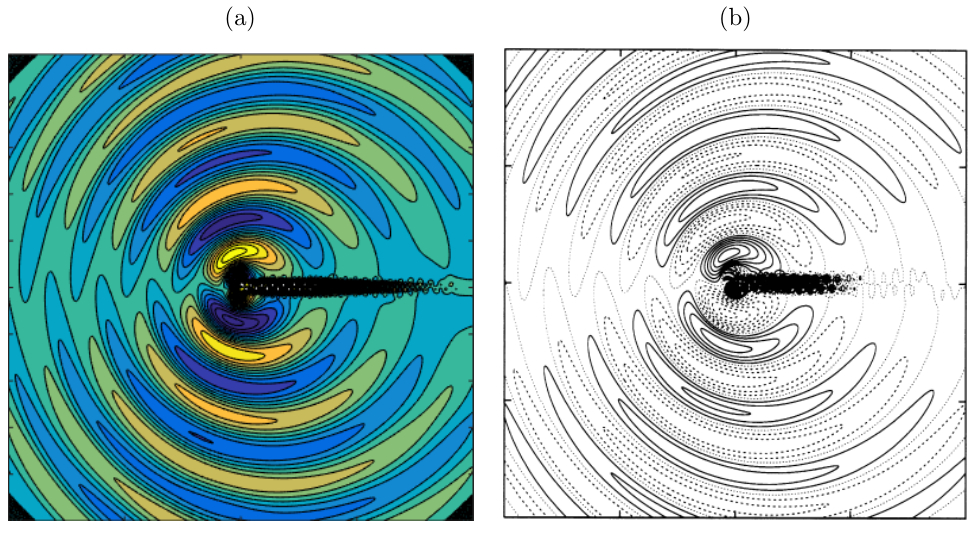}
      \caption*{Figure 4.(Color online) Pressure fluctuation in the physical domain for (a) the present model and for (b) the DNS study of Inoue and Hatakeyama.\citep{Inoue2002}
         The size of the domain ($-100<x<100$, $-100<y<100$) as well as the contour levels ($\Delta{\tilde{p}}_{min}=\pm 0.1{\mathrm{Ma}}^{2.5}$ with an increment of $0.0025{\mathrm{Ma}}^{2.5}$)
         for (a) and (b) are the same.}
%      \label{cylinderFarField}  
    \end{figure}
    Simulation results for the pressure fluctuations are remarkably similar to those obtained by the DNS study.
    As expected, the wave field  has a dipole character with a slight asymmetry due to the doppler effect. 
    In particular, $(\Delta{\tilde{p}})_{rms}$ presents largest values for an angle $\theta = 78.2^\circ$ in good agreement
    with Inoue and Hatakeyama\cite{Inoue2002}($\theta = 78.5^\circ$).
    Finally, we test the wave propagation showing the decay of the fluctuation pressure peaks (fig. 5).
    In particular, we consider the pressure fluctuations along the radial distance in the direction $\theta = 90^\circ$ 
    for three time steps, that is $\Delta{\tilde{p}}_{90} \equiv \Delta{\tilde{p}}(r, 90, t_{1,2,3})$ 
    (fig. 5a).
    As predicted by the theory,\citep{Phillips1956} the peaks of the wave forms $\Delta{\tilde{p}}_{90}$
    follow the decay $r^{-1/2}$ (fig. 5b). 
    \begin{figure}[t]  % 5
        \centering
         \includegraphics[width=0.6\textwidth]{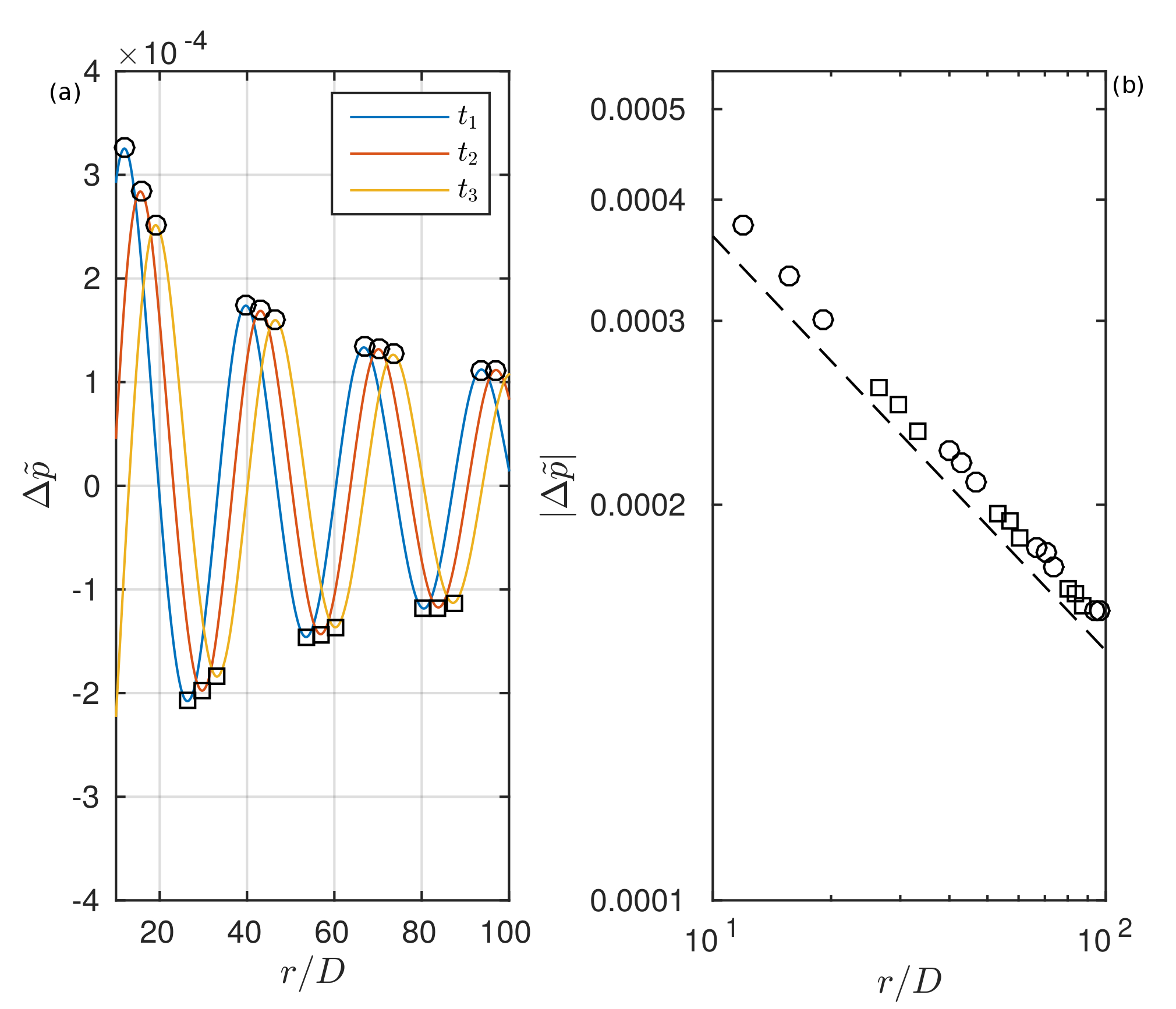}
         \caption*{Figure 5.(Color online) Propagation and decay of pressure waves ($\Delta{\tilde{p}}(r,90^\circ)$).
          (a) Waves for three time steps: $t=t_{1}$, $t_{2}=t_{1}+\Delta{t}$, $t_{3}=t_{1}+(2*\Delta{t})$, with $\Delta{t}=0.7$.
          (b) Decay of positive ($\circ$) and negative ($\Box$) wave peaks
          and $r^{-1/2}$ theoretical decay ($---$).}
%          \label{waveDecay}
    \end{figure}

\section{COMPUTATION OF THE NOISE RADIATED BY A TURBULENT JET}

    The subtle mechanisms that generate sound from turbulent jets are not yet well understood
    and widely debated (e.g Jordan And Colonius,\cite{Jordan2013} Jordan and Gervais,\cite{Jordan2008}
    Kearney-Fischer \cite{Kearney2013}). 
    However, the noise produced by turbulent jets is well characterized experimentally. This fact provides
    the possibility to test numerical models that can help to better understand sound production processes. 
    In the last twenty years,  numerous numerical investigations have been conducted for the turbulent
    jet over a wide range of Mach and Reynolds numbers (e.g. Cavalieri \textit{et al.},\cite{Cavalieri2011} 
    Bogey \textit{et al.},\cite{Bogey2012} Nichols and Lele\cite{Nichols2012}). 
    In the following, we will mainly refer to the case study proposed by Lew \textit{et al.}\cite{Lew2010}
    who compared a commercial software based on LBM (PowerFLOW)
    jet noise predictions with the ones of a Navier stokes solver\citep{Uzun2004} (referred as NS-LES in what follows) and
    experimental data.\citep{Bridges2003}$^,$\citep{Tanna1977}
    Here, this comparison is proposed again adding simulation results from the present LBM model.
    For seek of brevity only the main test results will be presented. 

    A subsonic turbulent jet exiting from a cylindrical pipe with a diameter $D_{j} = 0.0508$ m
    ($R_j$ the radius) and length $L = 0.508$ m is considered. 
    We imposed a constant velocity profile at the pipe inlet with $Ma = 0.3$ and a fluid viscosity
    $\nu = 0.0011$ $m^{2}/s$. As a result of the development of the boundary layer,
    an initially laminar jet exits the pipe with $\mathrm{Ma} \sim 0.4$ and $\mathrm{Re} \sim 6000$,
    where $\mathrm{Ma} = U_{j}/c_\infty$ and $\mathrm{Re} = U_{j}D_{j}/\nu$.
    In particular, the jet velocity at the pipe exit  ($U_{j}$) is $128$ $m/s$.
    The pipe diameter and the jet exit velocity are chosen to be consistent with Tanna's experiments
    (set point 2 in Tanna's experimental test matrix\citep{Tanna1977}).
    The computational domain of $[40\times40\times40]\times D_j$ is divided into five refinement levels 
    for around $80\times10^{6}$ total number of grid points (fig. 6).
    \begin{figure}[h]  % 6
        \centering
         \includegraphics[width=0.6\textwidth]{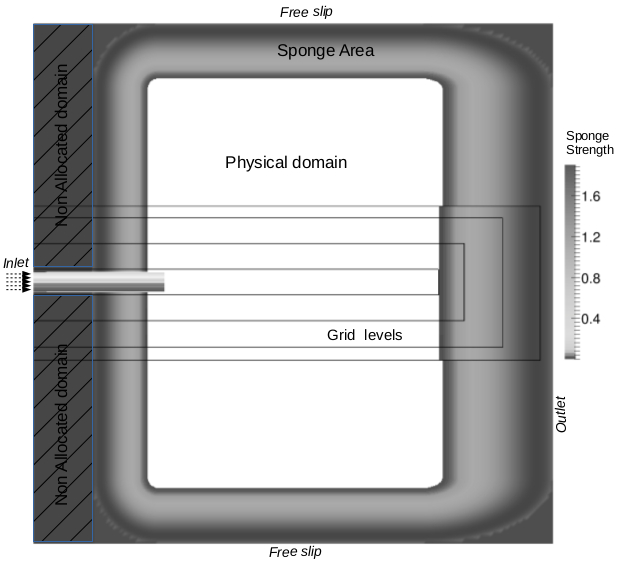}
         \caption*{Figure 6. Simulation setup for the aeroacoustic computation of a turbulent jet ($\mathrm{Ma} = 0.4, \mathrm{Re} = 6000$).
         In order to optimize the computational cost, the domain on the sides of the pipe is not considered in the calculation.}
%         \label{jetSetup}
    \end{figure}
    Free slip boundary condition and costant pressure conditions are enforced on the lateral
    boundaries and the right boundary respectively (fig. 6).
    The absorbing dynamics is imposed near all boundaries and for
    the outflow region, the jet similarity solution\citep{Pope2000} is used as a reference field.
    The inclusion of the pipe geometry in the computational domain, performed with a simple bounce back
    boundary condition,\citep{Gallivan1997} induces the spontaneous laminar-turbulent transition of the jet
    without the need of artificial forcing techniques, that are well known  sources of spurious acoustic
    noise.\cite{Bogey2005}${}^,$\cite{Lew2010}${}^,$\cite{Lew2004}
    Although the pipe geometry is placed inside the finest refinement level 
    ($\delta{x}= 8\times10^{-4}$ m, $dt= 1.4\times10^{-6}$ s), the boundary layer remains under 
    resolved by at least one order of magnitude without the use of a wall model. 
    However, this resolution is enough to create the conditions, at the exit of the pipe, for the
    jet shear layer to change from laminar to turbulent.  
    Finally, as it has been done by Lew \textit{et al.},\citep{Lew2010} we performed an under-resolved DNS
    since we do not use any explicit turbulence model. 
    
    The simulation ran for about 44308 iterations of the coarse level equivalent to ${1.00}$ s.
    The statistics were computed after $0.1$s ($0.3$s for the second order statistics), for which
    the simulation reached a steady state in the statistical sense.
    The computation lasted for 3 days of runtime on 400 processors (Sandy Bridge Intel(R) Xeon(R)
    CPU E5-2660 @ 2.20 GHz).

    The structure of the flow field is first presented by the vorticity (fig. 7).
    \begin{figure}[h]  % 7
        \centering
         \includegraphics[width=0.5\textwidth]{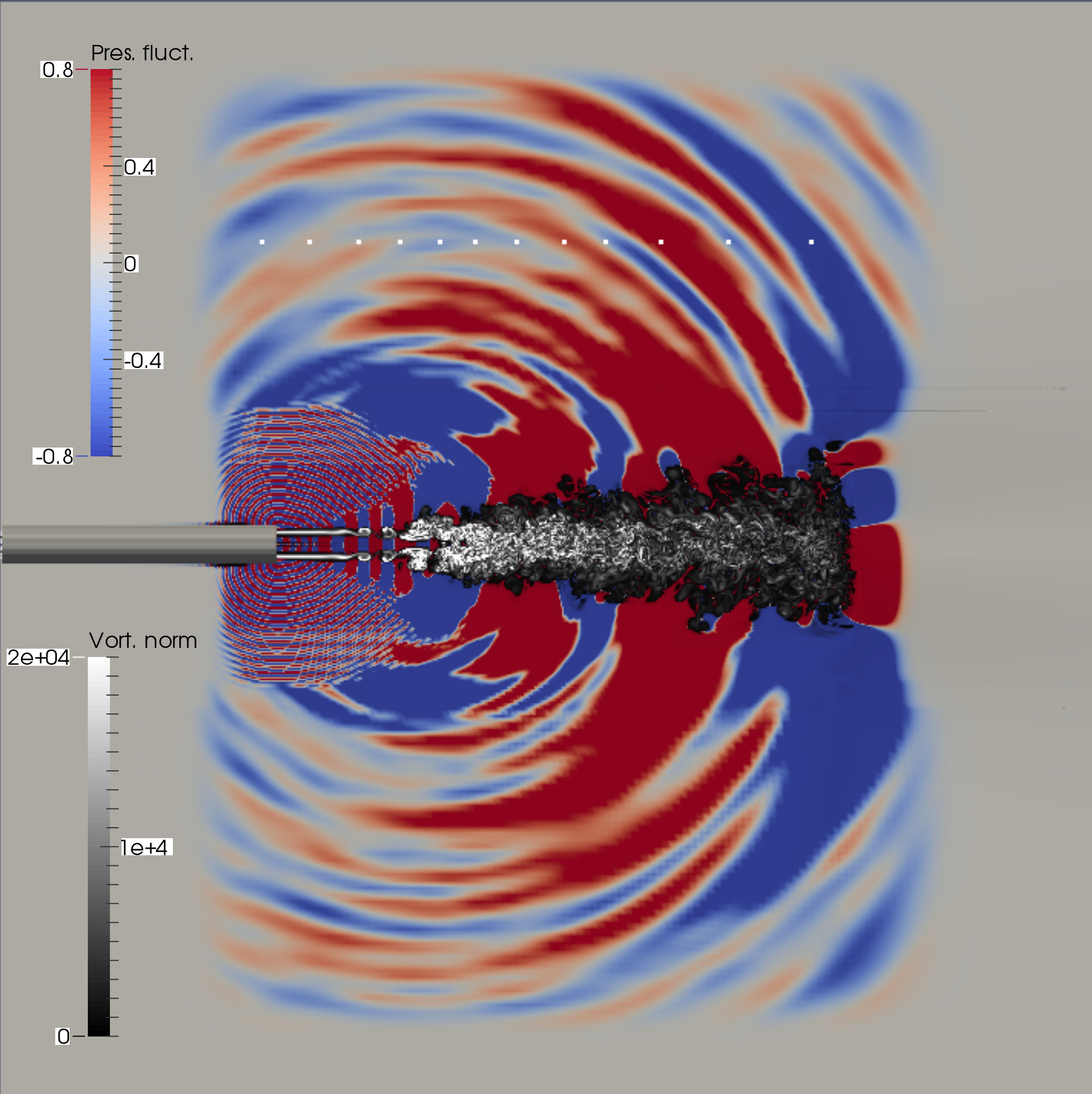}
         \caption*{ Figure 7.(Color online) Snapshot for the vorticity norm and acoustic field ($\Delta{\tilde{p}}$) for the turbulent jet ($\mathrm{Ma} = 0.4, \mathrm{Re} = 6000$).
           The positions of the acoustic probes are also shown (white points).}
%             \label{vortAndAc}
    \end{figure}
    At the pipe exit  instability waves (like Kelvin-Helmholtz),
    grow in the annular shear layer and develop large coherent structures (vortex) which cause the
    mixing of the jet and the ambient fluid to start. The complete transition towards a turbulent flow
    happens only where the large eddies merge and interact causing the potential core of the jet
    to break up.
    From this point the centerline velocity decays rapidly with increasing axial distance from the inlet
    and the jet spreads laterally with a constant spreading
    rate. In fig. 8 we report the decay for the centerline of the mean axial velocity
    ($U_c(x)$).   
    \begin{figure}[h]   % 8
        \centering
         \includegraphics[width=0.6\textwidth]{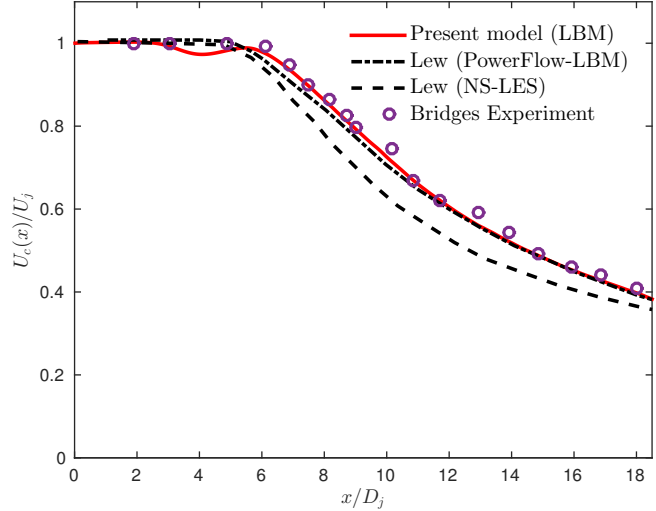}
         \caption*{Figure 8.(Color online) Mean streamwise velocity profile along the jet centerline axis ($\mathrm{Ma} = 0.4, \mathrm{Re} = 6000$) for the present model compared
           with the profiles reported for PowerFLOW,\cite{Lew2010} NS-LES\cite{Lew2010} and experimental data for $\mathrm{Ma} = 0.5$.\cite{Bridges2003}}
%       \label{velDecay}
    \end{figure}
    The first part of the centerline profile is close to the experimental values
    (for $\mathrm{Ma} = 0.5$\citep{Bridges2003}) and to those of both PowerFLOW and NS-LES.
    The potential core length, $l_p=7D_j$, defined as the axial distance from the inflow boundary where
    $U_{c}(l_p) = 0.95U_{j}$,  result to be in the experimental range $6/7D_j$ but
    slightly longer with respect to PowerFLOW and NS-LES ($6D_j$). For bigger axial distances the profile
    tend to overlap the PowerFLOW profile.   
   The centerline profile for the mean axial turbulence intensity $u_{c}(x)_{rms}$ is reported in 
   fig. 9.
   \begin{figure}[ht]   % 9
        \centering
          \includegraphics[width=0.6\textwidth]{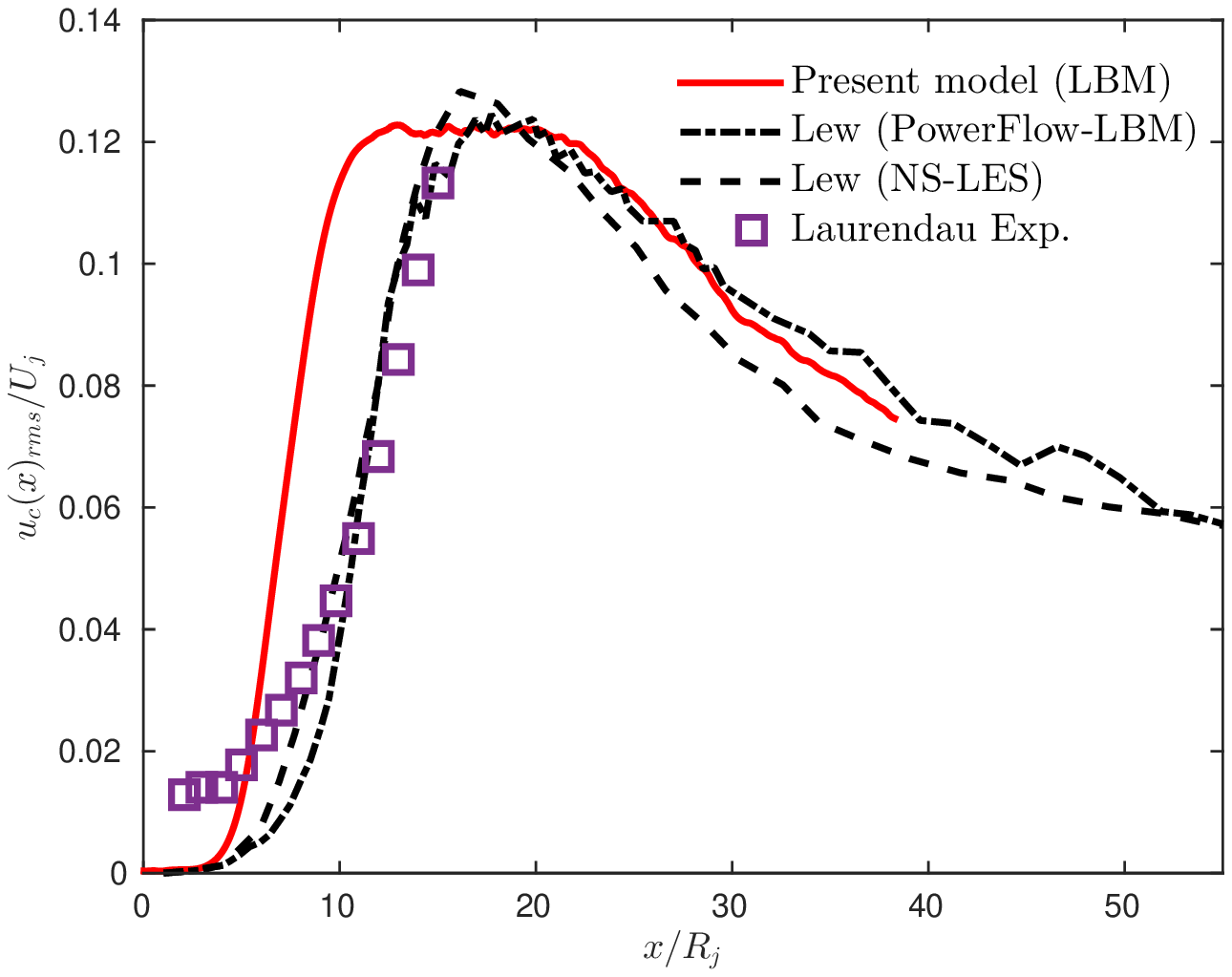}
         \caption*{Figure 9.(Color online) RMS streamwise velocity profile along the jet axis  ($\mathrm{Ma} = 0.4, \mathrm{Re} = 6000$) for the present model compared
           with the profiles reported for PowerFLOW,\cite{Lew2010} NS-LES\cite{Lew2010} and experimental data for $\mathrm{Ma} = 0.3$.\cite{Laurendeau2006}}
%                  \label{rmsCenterline}
   \end{figure}
   The $u_{c}(x)_{rms}$ peak value and its following decay are in good agreement
   with the profiles reported for PowerFLOW and NS-LES. However, we remark that the sudden increase
   in $u_{c}(x)_{rms}$ starts earlier than expected. Further investigations are needed to
   better understand the origin of this inaccuracy.
   
  Additional test results for the turbulent flow field including, self-similar profiles for both mean
   velocity and Reynolds stresses fields, spreading rate and velocity and pressure spectra have
   been reported by Malaspinas.\citep{Malaspinas2015} Moreover, in this work one can find 
    an interesting comparison with the results obtained using standard LBM methodologies.   
   
   Results for the acoustic field ($\Delta{\tilde{p}}$) are presented in fig. 7.
   It is well known that most of the radiated sound from turbulent jets is generated just after
   the potential core break up. As expected, acoustic waves shown in fig. 7 are clearly
   coming from this region. However, a more careful look reveals also the
   presence of a distinct source near the pipe exit which generate a lobated high frequency noise.
   The latter is most probably generated at the fluid-solid interface due to the presence
   of unresolved scales or unadapted boundary conditions (bounce-back) for the pipe geometry. In future,
   more sophisticated boundary conditions will be used to investigate the origin of this noise.
   However, these spurious effects are rapidly dissipated by the coarsening of the grid and do not affect significantly
   the accuracy of the acoustic spectra. 
   For the tests on the acoustic field,
   virtual pressure probes are placed sufficiently far away from the turbulent region of the jet.
   In particular, the acoustics probes are placed at distance of $11{D_j}$ from the
   centerline jet axis (fig. 7). For comparison with the far field experimental\citep{Tanna1977}
   and numerical data,\cite{Lew2010} we extrapolated acoustic probes data to the far field using the $1/r$
   correction from the source region located on the jet centerline axis
   at $x_{s} = 7 D_j$ ($ p_{farField} = \frac{r_{probe}}{r_{farField}}\cdot p_{probe}$).
   Although, this extrapolation does not take into account the Doppler effect, it has been shown to be
   a good first approximation.\citep{Bogey2003}${}^,$\citep{Lew2010} 
   The far field is extrapolated for on arc at a radial distance of $R = 72\cdot D_j$ from the center
   of the jet inflow boundary. It is worth to emphasize that distances $r$ used in the correction 
   are measured with
   respect to the source point on the centerline at $x_{s}$ whereas the far field distance $R$
   is referred to jet center. 
   In fig. 10 and 11 are reported $1/3$ octave spectra for two different
   angles ($\theta_{j}=45^{\circ}$ and $\theta_{j}=75^{\circ}$) measured relative to the jet centerline axis.
   \begin{figure}  % 10	 
       \centering
        \includegraphics[width=0.6\linewidth]{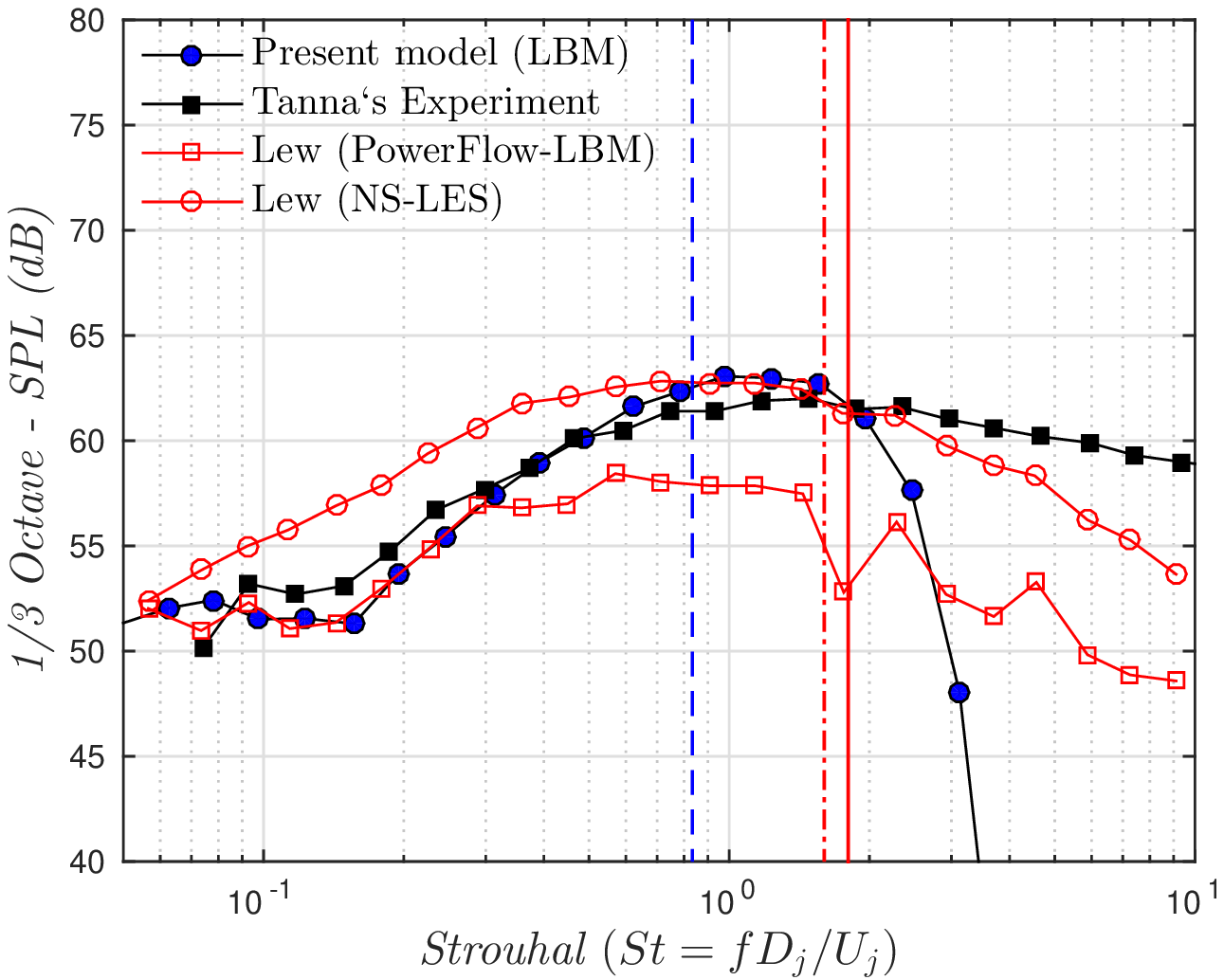}                      
        \caption*{Figure 10.(Color online) 1/3-Octave spectra  for $\theta_{j}=75^\circ$ 
        compared with the one reported for PowerFLOW,\cite{Lew2009} NS-LES\cite{Lew2010} and Tanna's experimental data.\citep{Tanna1977}
        Vertical lines mark the strouhal cut-off for the different numerical models: '--' for the present model, '-.-' for PowerFLOW and '-' for NS-LES.}
%          \label{octaveSpectra75}
   \end{figure}
   \begin{figure}[h]  % 11
        \centering
         \includegraphics[width=0.6\linewidth]{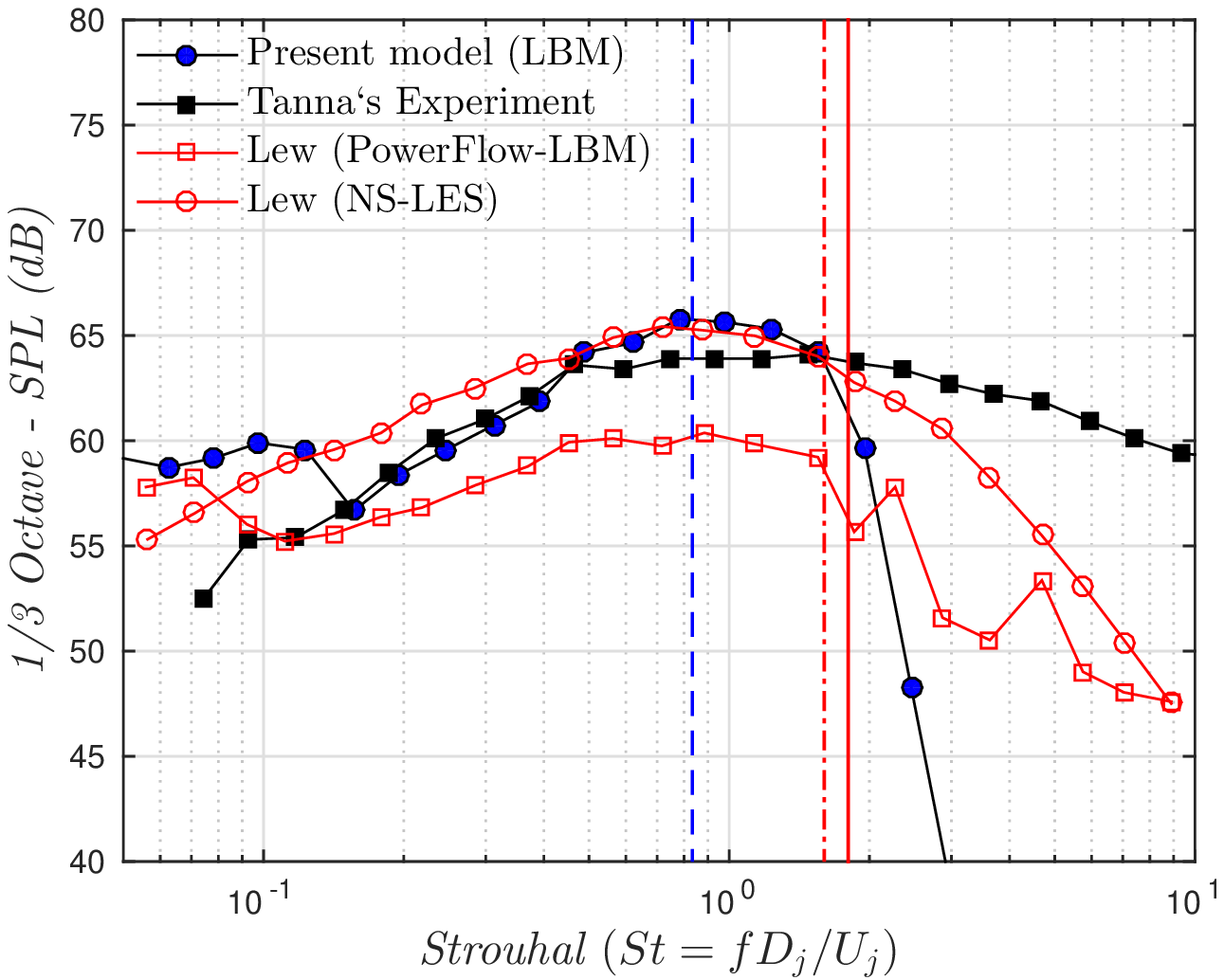}                        
        \caption*{Figure 11.(Color online) 1/3-Octave spectra  for $\theta_{j}=45^\circ$ 
        compared with the one reported for PowerFLOW,\cite{Lew2010} NS-LES\cite{Lew2010} and Tanna's experimental data.\citep{Tanna1977}
        Vertical lines mark the strouhal cut-off for the different numerical models: '--' for the present model, '-.-' for PowerFLOW and '-' for NS-LES.}
%          \label{octaveSpectra45}
   \end{figure}
   Following Lew \textit{et al.},\cite{Lew2010} the maximum resolvable frequency is calculated as
   $f_{max} = c_{\infty}/\lambda_{max}$ where $\lambda_{max} = 12\cdot \delta{x}$.
   Since $\delta{x} = 0.0127 $ at the refinement level where the acoustic probe are located,
   $St_{max}\sim0.8$, with $St_{max} = f_{max}D_{j}/U_j$.  
   The  over all level and the shape of the spectrum for both the angles are in agreement
   within a range of $+/- 1$ dB with experiments. Compared to the other numerical simulations,
   both NS-LES and PowerFLOW our model seems to be more accurate for $0.15<St<2$. However, the spectral
   shapes for small angles, that correspond 
   to probes near to the outflow boundary, suffer by an overestimation (maximum 5 dB) in the
   low-frequency part ($St < 0.15$).
   This low frequency is probably produced by the interaction of the turbulent flow or the outgoing 
   wave field with outlet sponge layer. A bigger sponge layer on the outflow boundary and/or
   the coupling of this with an hyper-viscosity layer may resolve this issue and will be
   investigated in future studies. Nonetheless, consistently with the result reported for PowerFLOW,
   the amplitude of the spectra for our simulation are not affected by the overestimation which characterizes
   instead the NS-LES spectra. This overestimation is probably due to the additional spurious noise generated
   by the artificial forcing used in the NS-LES simulation.\cite{Lew2010} Let us note that the porous
   Ffowcs Williams Hawkings surface integral acoustic method have been used to compute
   the far-field noise for NS-LES, instead of using the more simple $1/r$ correction.
   Compared to PowerFLOW instead, the present model shows a more rapid attenuation of the amplitudes after the
   cut off, probably  due to the coarser $\delta{t}$, but better values near the cut-off line. 
   
   \begin{figure}   % 12
        \centering
         \includegraphics[width=0.6\textwidth]{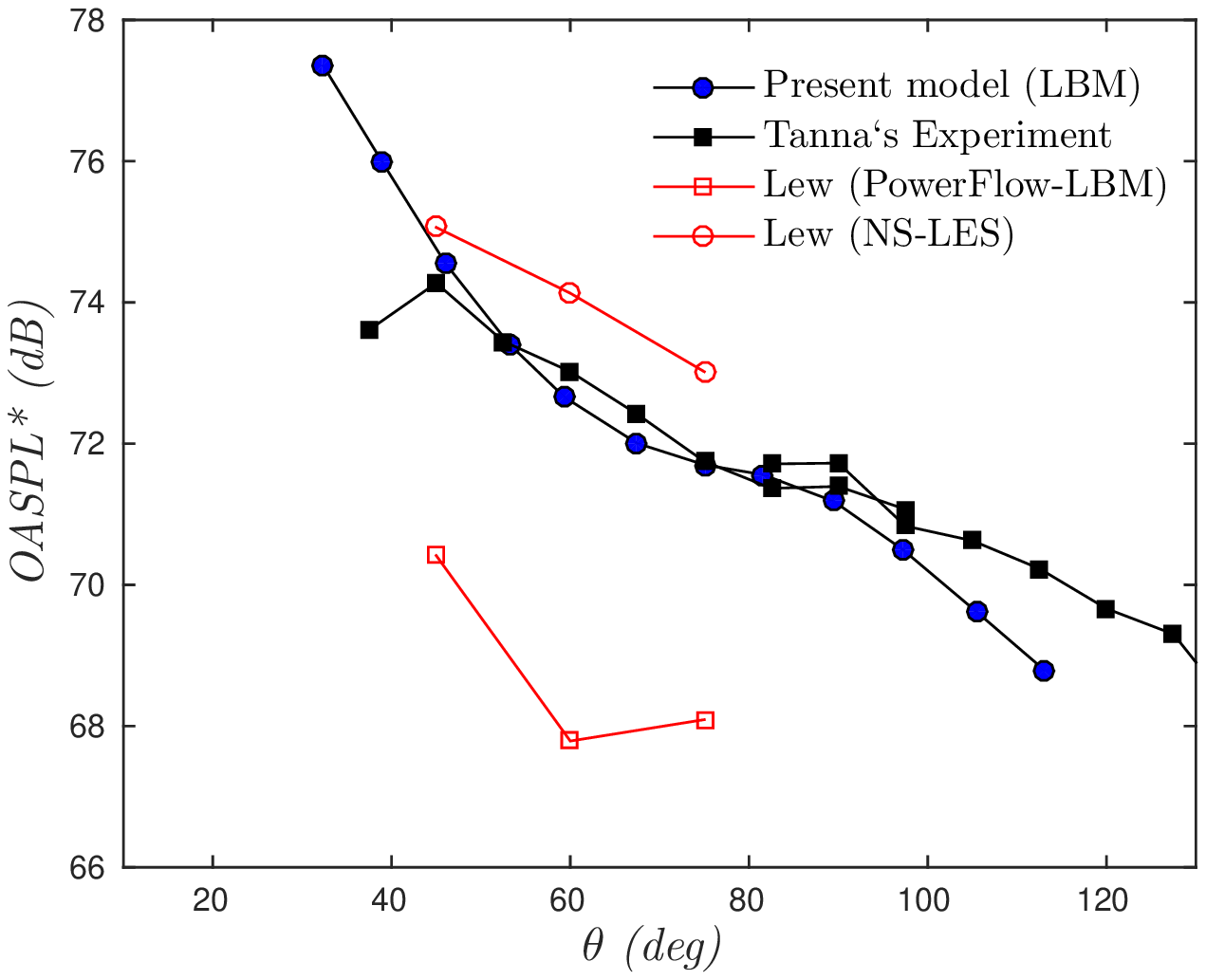}
         \caption*{Figure 12.(Color online) Overall sound pressure level (OASPL*) directivity at the far field distance of 72D as a
                      function of the observation angle $\theta_j$.
                      The computed OASPL* is computed from the 1/3 octave band spectra using a limited band test range (0.07-3 $St$).
                      For comparison, the OASPL* for Tanna's experimental data\citep{Tanna1977} and for PowerFLOW and NS-LES numerical
                      data\cite{Lew2010}$^,$\cite{Lew2009} are also shown.}
%                  \label{OASPL}
   \end{figure}
   Finally, we present test results for the directivity of the acoustic field.
   Fig. 12 shows the Over All Sound Pressure Level (OASPL*) as function of $\theta_j$ computed,
   for model testing purposes, from the 1/3 octave band spectra in the limited band range (0.07-3 $St$)
   for which we have data for the present model (fig. 10, 11). 
   As done before, the OASPL* for Tanna's data \citep{Tanna1977} is used for comparison as well as the ones for 
   PowerFLOW and NS-LES. Regarding the latter two 1/3 octave spectra were found for three angles 
   \cite{Lew2010}$^,$\cite{Lew2009} and for the Tanna's experiment 1/3 octave data were not available
   for $\theta_{j}<37.5^{\circ}$.
   As expected, for small angles  the overestimation of the present model as already mentioned above is due to the presence of a
   spurious low frequency in the octave spectra for the probes near to the outflow region.
   However, in the range of angles $40^{\circ}<\theta_{j}<120^{\circ}$, the results are remarkably close to the 
   experimental values providing us the evidence of the accuracy of the present model for jet aeroacoustics 
   even when compared with other numerical models.
    Let us finally note that the numerical data used for comparison to PowerFLOW\citep{Lew2010} 
   are less accurate than the more recent ones reported by Lew \textit{et al.}\cite{Lew2014} for a more
   sophisticated numerical simulation (realistic nozzle geometry, Ffowcs Williams-Hawkings method for the acoustic far field,
   non standard LBM for high speed flows etc ...).

\vspace{1em}
  \section{DISCUSSION AND CONCLUSIONS}

  In this paper we presented in great detail all the building blocks of a lattice Boltzmann method 
  simulation of aeroacoustics. The complete numerical model is built in the 
  open source library PALABOS. 
  A modified collision model improves the accuracy and the stability 
  of the standard BGK-LBM (or MRT-LBM) thanks to a recursive regularized LBM dynamics. In addition,
  the model makes use of an optimal absorbing strategy as NRBC and includes a multi-domain grid
  refinement which allows us to deals with complex multi-scale fluid flows.
  This work represents the first attempt to combine these components together and 
  test them on complex aeroacoustic applications using both numerical and experimental data.

  First, a laminar 2D model benchmark for the sound generated by cylinder in an uniform flow
  ($Ma=0.2$, $Re=150$),
  provides the evidence that the model correctly recovers both the sound source mechanism and the wave propagation.
  The wave field turns out to be remarkably similar to the one obtained in a numerical (DNS)
  investigation,\cite{Inoue2002} showing the same directivity. The wave amplitude decay follows
  the $r^{-1/2}$ theoretical law.

  The model performance was then tested on a more complex 3D turbulent jet noise computation
  ($Ma=0.4$, $Re=6000$). The results for this simulation have been compared with experimental
  and numerical data from both a  Navier-Stokes solver (NS-LES) and the LBM PowerFlow closed
  source software (PowerFLOW).
  Overall, results for the flow field and the acoustic field from the present model are
  in good agreement with experiments and predictions from NS-LES and PowerFLOW. 
  Although the acoustic predictions are affected by a spurious low frequency, generated
  near the outflow region, they are within  $+/-1$ dB from the experimental data for
  $0.15<St<2$. As reported for PowerFLOW, thanks to the inclusion of the pipe geometry in the computational domain, 
  the acoustic spectra  are not affected by the overestimation which characterizes the the NS-LES spectra.
  Compared to PowerFLOW, our model seems to be more accurate near the cut-off frequency.

  In conclusion, this paper establishes an LBM optimized computation model which has been
  shown to be accurate for aeroacoustic computation. In particular, it is able to accurately
  reproduce the physical mechanisms responsible for sound generation and its propagation over
  large distances. 
  Therefore, it may be used, for example, to perform direct aeroacoustic computations in order
  to study the noise generation mechanisms and develop noise reduction strategies
  for terrestrial vehicles.  
  In particular, a very promising application could be a coupling 
  with dedicated noise identification algorithms\cite{Vergnault2013} for the identification 
  of sound sources. Finally, the use of optimization techniques, such as adjoint methods,
  \cite{Vergnault2014} could help in the design of low noise emission vehicles. 

  Currently, the model is limited to low Mach number computations for athermal flows and
  remains stable, when tested on the 3D turbulent jet, for $\mathrm{Ma} \leq 0.6$.
  The model extension to fully compressible flows, needed in aerospace, or geological applications,
  is currently under investigations with the use of multi-speed models.\cite{Nie2008}

 \medskip

 \noindent \textbf{ACKNOWLEDGEMENTS}
 
 \setlength{\parindent}{0.7cm} 
  
    This work was supported by the Swiss National Science Foundation SNF (project N. $200021$\_$137942$)
    and the PASC project (http://www.pasc-ch.org/). 
    We would like to thank the University of Geneva (Geneva, Switzerland) and the CADMOS project
    (http://www.cadmos.org/) for providing the computational resources.
    F. Brogi thankfully acknowledges Jonas Latt and Yann Thorimbert for the enlighting discussions.


\begin{thebibliography}{99}
\bibitem{DdQq}
 DdQq notation: d is the number of dimensions and q refers to the number
 of quadrature points which corresponds to the number of lattice velocities.
\bibitem{DampingTarget} 
  The damping target is defined as the amplitude ratio of the reflected and the incident disturbances:
  $\eta\equiv20\log_{10}\frac{|A_{reflected}|}{|A_{incident}|}$.
\end{thebibliography}
 \end{document}